\documentclass[twocolumn,amsmath,amssymb]{revtex4}

\usepackage{graphicx}
\usepackage{bm}
\usepackage{dcolumn}
\usepackage{color}
\usepackage{float}

\newcommand{\be}{\begin{equation}}
\newcommand{\ee}{\end{equation}}
\newcommand{\bea}{\begin{eqnarray}}
\newcommand{\eea}{\end{eqnarray}}
\newcommand{\non}{\nonumber}

\begin{document}

\title{The entropy of an acoustic black hole in Bose-Einstein condensates}

\author{Massimiliano Rinaldi}
\email{rinaldi.max@gmail.com}
\affiliation{Istituto Nazionale di Fisica Nucleare \\
 Sezione di Bologna, \\ Via Irnerio 46, I-40126 Bologna, Italy.}

\begin{abstract}

\noindent We compute the entanglement entropy associated to the Hawking emission of a $(1+1)$-dimensional acoustic black hole in a Bose-Einstein condensate. We use the brick wall model proposed by 't Hooft, adapted to the momentum space, in order to tackle the case when high frequency dispersion is taken in account. As expected, we find that in the hydrodynamic limit the entropy only depends on the size of the box in the near-horizon region, as for gravitational $(1+1)$-dimensional black holes. When dispersion effects are considered, we find a correction that depends on the square of the size of the near-horizon region measured in units of healing length, very similar to the universal correction to the entropy found in the case of spin-1/2 Heisenberg XX chains. 
\end{abstract}


\maketitle

%%%%%%%%%

\section{Introduction}

%%%%%%%%%

\noindent When Hawking realized that a Schwarzschild black hole emits radiation like a  black body with a temperature determined by its mass \cite{Hawking}, investigations focused on the connection between the Hawking-Bekestein formula \cite{bek} for the entropy $S={1\over 4}M_{\rm Pl}^{2}A$, where $A$ is the area of the horizon, and some consistent microscopic counting of degrees of freedom. Generally speaking, there is a  large consensus on defining the entropy by using the von Neumann formula $S=-{\rm Tr} (\rho\ln \rho)$ where $\rho$ is a density matrix. In particular, one can associate the density matrix to the sub-state formed by the outside region of the black hole. In this case, the entropy measures the degree of entanglement between the modes in the two sides of the horizon \cite{entangl}. Alternatively, entropy can be defined through the statistical mechanics of a system in the vicinity of the horizon \cite{GibbHawk,thooft}. Remarkably, the two characterizations coincide and agree with the Bekenstein-Hawking formula up to the factor 1/4, whose origin is still unknown. It should also be mentioned that these microscopic realizations of the entropy suffer from ultraviolet divergences, that can be cured by introducing a cutoff at around the Planck scale, see \cite{brustein} for recent developments and \cite{reviewentropy} for a review.

The lack of experimental evidence for Hawking radiation  is mainly due to the smallness of  $\hbar$ and $c^{-1}$. However, as first noticed by W. Unruh in 1981, in condensed matter physics there are systems that closely mimic curved spacetime configurations, and where the speed of light is effectively replaced by the speed of sound waves, so the suppression of quantum effects can be lifted by several order of magnitudes \cite{unruh}. In particular, an irrotational fluid flowing through a device able to accelerate it to supersonic speed can generate a thermal flux of  phonons  that shows the same characteristics of the Hawking radiation emitted by a black holes, see \cite{cimento} and references therein. This possibility was studied in the context of Bose-Einstein condensates (BEC) and many other systems, see e.g.  \cite{reviewbec}. Although no analog formulae to the Hawking-Bekenstein one are known for these dumb holes, we expect that  entropy can be associated to the phonons created via the Hawking mechanism. In fact, the acoustic horizon acts as a partitioning screen, which is a sufficient condition to create entanglement and, therefore, entanglement entropy.

In this paper, we would like to address the calculation of the entanglement entropy associated to the analog of the Hawking radiation created in  (1+1)-dimensional acoustic black holes in dilute BEC gas. In order to avoid typical infrared divergences occurring in this kind of bi-partite systems, we confine the region in which we compute the entropy into a box of size $L$ and located at an arbitrary distance $\epsilon$ near the horizon. The (1+1)-dimensional acoustic black holes was intensively studied both analytically \cite{balb-carus-fabbri,mayo-fabbri,mayo-fabbri-rinaldi,analyticBEC} and numerically \cite{numericBEC} as it might be experimentally realizable.  In the limit where the wavelength of the modes are much larger than the healing length of the gas, one can neglect the high frequency dispersion typical of this system (the so-called hydrodynamic limit). In this case, we expect that the entropy is proportional, at the leading term,  to  $\ln (L/\epsilon)$. This is due to the fact that the mode equation in $(1+1)$ dimensions is nearly conformally invariant, exactly like in the case of a $(1+1)$-dimensional gravitational black hole \footnote{In the case of gravitational (1+1)-dimensional black holes obtained by dimensional reduction, conformal invariance is obtained by neglecting a potential term, see e.g. \cite{balb-carus-fabbri}.}. Therefore, the entropy is purely ``geometric'' and arbitrary, in the sense that it cannot depend on the parameters of the black hole. In this paper, we verify  this  by employing the brick wall model proposed by 't Hooft in \cite{thooft}, see also \cite{indians}. We stress that the entanglement entropy computed here refers only to the phononic radiation produced by the Hawking mechanism, and has nothing to do with the thermodynamic entropy of the Bose-Einstein condensate, which vanishes. 

The main results of our work concern however the case when dispersion is taken in account, and conformal invariance is broken. In fact, the dispersion typically introduces a high-order differential operator in the mode equations, with a prefactor that depends on the healing length, in analogy with certain gravitational models endowed with modified dispersion relations, see e.\ g.\ \cite{mdr}. Because of this term, the entropy is no longer arbitrary and it is reasonable to expect that it depends on the healing length. Indeed, we find that this is the case, by using the brick wall technique introduced by 't Hooft and adapted to the momentum space. As far as we know, this is the first time that such a method is employed in the context of acoustic black holes, and we find that our adaptation to momentum space becomes a powerful tool to compute the entanglement entropy, especially when high frequency dispersion is present.

The plan of the paper is the following. In the next section we briefly review the equations governing the (1+1)-dimensional BEC in both the hydrodynamic limit and the dispersive case. In Sec. III we recall the brick wall model, and we show how one can use it in momentum space. In Sec. IV we apply the method to compute the entanglement entropy in the acoustic black hole with dispersion. In Sec. V we discuss our results, and compare them with some numerical calculations.

%%%%%%%%%%
\section{The setup}
%%%%%%%%%

In the dilute gas approximation \cite{stringari}, the BEC can be described by an operator $\hat \Psi$ that obeys the equation
\bea
i\hbar \partial_{t}\hat\Psi=\left( -{\hbar^{2}\over 2m} \vec{\nabla}^{2}+V_{\rm ext} + g\hat\Psi^{\dagger}\hat\Psi\right)\hat\Psi\ ,
\eea
where $m$ is the mass of the atoms, $g$ is the non-linear atom-atom interaction constant, and $V_{\rm ext}$ is the external trapping potential. The wave operator satisfies the canonical commutation relations $[\hat\Psi(t,\vec{x}),\hat\Psi(t,\vec{x}')]=\delta^{3}(\vec{x}-\vec{x}')$. To study linear fluctuations, one substitutes $\hat\Psi$ with $\Psi_{0}(1+\hat\phi)$ so that $\Psi_{0}$ satisfies the Gross-Pitaevski equation 
\bea\label{GP}
i\hbar \partial_{t}\Psi_{0}=\left( -{\hbar^{2}\over 2m} \vec{\nabla}^{2}+V_{\rm ext} + gn\right)\Psi_{0}\ ,
\eea
and the fluctuation $\hat\phi$ is governed by the Bogolubov-de Gennes equation
\bea\label{bdg}
i\hbar \partial_{t}\hat\phi=-{\hbar^{2}\over 2m}\left(\vec{\nabla}^{2}+2{\vec{\nabla}\Psi_{0}\over \Psi_{0}}\vec{\nabla}\right)\hat\phi+mc^{2}(\hat\phi+\hat\phi^{\dagger}),
\eea
where $c=\sqrt{gn/m}$ is the speed of sound, and $n=|\Psi_{0}|^{2}$ is the number density. We now focus on the $(1+1)-$dimensional case and consider a configuration with constant $v$ and $n$, while the speed of sound $c(x)$ smoothly decreases from the subsonic region to the supersonic one, and it is equal to $v$ at $x=0$, as in \cite{balb-carus-fabbri}.

This is possible provided one modulates $g$, and hence the speed of sound $c$, by keeping the combination $gn+V_{\rm ext}$ unchanged \cite{numericBEC}. In this way, Eq.\ (\ref{GP}) admits the plane-wave solution $\Psi_{0}=\sqrt{n}\exp (ik_{0}x-i\omega_{0}t)$ where $v=\hbar k_{0}/m$ is the condensate velocity. To study the Bogolubov-de Gennes equation, we expand the field operator as $
\hat\phi(t,x)=\sum_{j}\left[\hat a_{j}\phi_{j}(t,x)+\hat a_{j}^{\dagger}\varphi_{j}^{*}(t,x)\right]\,$
and we find that the modes $\phi_j (t,x)$ and
$\varphi_j(t,x)$ satisfy the coupled differential equations   \cite{mayo-fabbri-rinaldi}
\bea\non\label{dispeqns}
\left[ i(\partial_t + v\partial_x) + \frac{\xi c}{2} \partial_x^2 -\frac{c}{\xi} \right] \phi_j &=& \frac{c}{\xi}
\varphi_j\ , \\
\left[ -i (\partial_t + v\partial_x) + \frac{\xi c}{2}\partial_x^2 - \frac{c}{\xi}\right] \varphi_j &=&\frac{c}{\xi} \phi_j  \ ,\label{cde}
\eea
 where $\xi=\hbar/(mc)$ is the healing length of the condensate. 
 
In these settings, the dispersive effects are signaled by the presence of $\xi$, which depends on the local velocity of sound and that is a non-perturbative parameter \cite{mayo-fabbri-rinaldi}. Therefore, to study the case when dispersion is negligible one must switch to the density-phase representation consisting in defining the density $\hat n_{1}$ and phase operators $\theta_{1}$ via
 \bea
 \hat\phi={\hat n_{1}\over 2n}+i{\hat\theta_{1}\over \hbar}\ ,
 \eea
 along the lines of \cite{balb-carus-fabbri}. With these definitions, the limit $\xi\rightarrow 0$ is well defined and one finds the single equation 
\begin{equation}\label{sde}
(\partial_t+v\partial_x)\frac{1}{c^2}(\partial_t+v\partial_x)\theta_{1} = \partial_x^2\theta_{1}\ .
\end{equation}
The analogy with gravitational black holes comes about when one notices that the above equation can be written as $\square \theta_{1}=0$, where the d'Alambertian is defined on the so-called acoustic metric
\bea\non\label{acmetric}
ds^{2}&\!=\!&{n\over mc}\left[-(c^{2}-v^{2})dt^{2}-2vdxdt+dx^{2}+dy^{2}+dz^{2}\right],\\
&&
\eea
and where one assumes that $\theta_{1}$ does not depend on the transverse coordinates $y$ and $z$ \cite{mayo-fabbri}. The metric shows an event horizon located where $c(x)=v$ and its structure is the same of the Painlev\'e-Gullstrand line element, up to the conformal factor $n/(mc)$.
Note that Eq.\ (\ref{sde}) is not conformally invariant unless $c$ is constant.

%%%%%%%%%
\section{The brick wall in momentum space}
%%%%%%%%%

We now turn the the calculation of the entropy with the brick wall method, originally described by 't Hooft in \cite{thooft}, where it was applied to a (3+1)-dimensional Schwarzschild black hole. The method is based on the counting of the modes of a massive scalar field, with Dirichlet boundary conditions, defined inside a box of size $L$ and placed at a distance $\epsilon$ from the horizon. The result is that the entropy is proportional to the area of the horizon. Also, the proper distance between the horizon and the edge of the box is a constant of the order of the Planck length. The physical interpretation of this entropy is discussed in great detail in \cite{Muko}.

For (1+1)-dimensional gravitational black holes, the result is radically different and, at the leading order, the entropy reads
\bea\label{entropy}
S\simeq{1\over 6}\ln \left({L\over \epsilon}\right)\ ,
\eea
in the limit $L/\epsilon\gg 1$ \cite{twodim}. In the case of the acoustic black hole studied here, the result is the same, at least at the leading order, if we do not account for high frequency dispersion. To show this, it is sufficient to assume that the modes that are solutions to Eq.\ (\ref{sde}) vanish at the boundaries of the segment $[\epsilon, L]$, where $\epsilon$ is located near the horizon, at $x=0$, and $L$ is the length of the near-horizon region, namely the region where we can linearize the speed of sound as $c(x)\simeq c(0)+\kappa x$ on each side of the horizon. In this expansion
\bea
\kappa={1\over 2v}{d\over dx}(c^{2}-v^{2})_{x=0}
\eea 
is the analog of the surface gravity of the black hole. The stationary solutions $f(x)$ of Eq.\ (\ref{sde}) can be found with a WKB approximation by writing the solution as
\bea\label{wkbsol}
\theta_{1}(x)={\theta_{0}\over \sqrt{f(x)}}\exp\left({i\over \hbar}\int^{x} f(x') dx'\right)\ .
\eea
This expression can be seen as the continuum limit of a function of the form $\exp (i\sum_{n} k_{n})$ for $n=0...n_{\rm max}$, where the sum counts the number of modes. Thus, in the continuum limit, the number of modes populating the interval $[\epsilon, L]$ with frequency $\omega$ is given by \cite{thooft}
\bea\label{xmodes}
n(\omega)={1\over \pi \hbar}\int_{\epsilon}^{L}f(x)dx\simeq{\omega\over \pi\kappa}\ln \left(L\over \epsilon\right) \ .
\eea
We now recall that the free energy and the entropy associated to massless spin-0 particles are given respectively by
\bea\label{freeen}
F=-\int_{0}^{\infty}{n(E)\over (e^{\beta E}-1)}dE\ ,\quad S=\beta^{2}{dF\over d\beta}\ ,
\eea
where $E=\hbar\omega$ and $\beta$ is the inverse of the temperature of the black hole,  $\beta=(k_{\rm B}T)^{-1}=2\pi/(\hbar\kappa)$. 

By replacing the expression (\ref{wkbsol}) into Eq.\ (\ref{sde}), and keeping only the leading terms in $\hbar$ we find that $f(x)\simeq \hbar\omega/( c(x)\pm v)$. By expanding around the horizon, where $c(x)\simeq v+\kappa x$, we see that the dominating solution for small $x$  is $f(x)\simeq \hbar\omega/\kappa x$, thus we  find Eq.\ (\ref{entropy}). This result confirms the validity of the brick wall model to compute the entanglement entropy when the acoustic metric has a horizon. In fact, the presence of a horizon is crucial to this result, as $f(x)\propto 1/x$ precisely because of the vanishing of the $g_{tt}$ term of the metric (\ref{acmetric}) at the horizon. We note further that  Eq.\ (\ref{entropy}) agrees also with the calculation of the leading term of entanglement entropy in bi-partite spin-chains \cite{cardy}, including the non-critical case, where conformal symmetry is only approximate \cite{cardy2}, as in the case studied here.
 
To tackle the dispersive case, it is convenient to derive this result also in momentum space. One defines the Fourier transform $\tilde\theta_{1}$ of the modes via
\bea
\theta_{1}(x)=\int {dp\over\sqrt{2\pi}}\,e^{ipx}\tilde\theta_{1}(p)\ ,
\eea
so that Eq.\ (\ref{sde}) in momentum space becomes
\bea\label{sdemom}
(\omega - vp){1\over \hat c^{2}}(\omega-vp)\tilde\theta_{1}=ip^{2}\tilde\theta_{1}\ ,
\eea
where, in the near-horizon approximation, 
 \bea
 {1\over \hat c^{2}} \simeq {1\over v^{2}}\left( 1-{2i\kappa\over v}\partial_{p}-{3\kappa^{2}\over v^{2}}\partial_{p}^{2}\right)\ . 
 \eea
The solutions $\tilde f(p)$ to Eq.\ (\ref{sdemom}) can be written in terms of Whittaker function. However, for our purposes it is sufficient to use again the WKB method by substituting in Eq.\ (\ref{sdemom}) the expression
\bea
\tilde\theta(p)={\tilde\theta_{0}\over\sqrt{\tilde f(p)}}\exp
\left({i\over\hbar}\int^{p} \tilde f(p')dp'\right)\ .
\eea
If we consider the large $vp/\kappa$ limit, i.e.\ we select wavelengths much smaller than the near-horizon region (whose size is approximately $v/\kappa$), and we recall that we are in the regime of linear dispersion $\omega=c(x) p$ , we find that, at the lowest order in the WKB expansion,
\bea\label{ndsoln}
\tilde f(p)\simeq {\hbar\omega\over \kappa p}\ .
\eea
As we are in momentum space, we now count the modes with an associated momentum between  $p_{\bf min}$ and $p_{\bf max}$. The first value correspond the the largest wavelength admitted in the near-horizon region and corresponds to the infrared contribution to the integral (\ref{xmodes}). The value $p_{\bf max}$ is interpreted as the minimal distance that we can probe with our modes. As we are considering the hydrodynamic limit, this implies $p_{\bf max}\xi\ll 1$. In analogy with Eq.\ (\ref{xmodes}), the number of modes is  defined as
\bea\label{numberND}
\tilde n(\omega)={1\over \pi \hbar}\int_{p_{\bf min}}^{p_{\bf max}}\tilde f(p)dp={\omega\over \pi\kappa}\ln \left(p_{\bf max}\over p_{\bf min}\right)\ .
\eea
By following the same steps as above, we find that
\bea
S={1\over 6} \ln \left(p_{\bf max}\over p_{\bf min}\right)\ .
\eea
This expression is equivalent to Eq.\ (\ref{entropy}) in terms of counting the numebr of degrees of freedom, however it stresses the fact that the entropy diverges in the ultraviolet non-locally, i.e. in no particular point in the near-horizon region. In this respect, the expression above reflects more closely the properties of the entanglement entropy \cite{brustein}.

Before considering the dispersive case, we recall that  the near-horizon region has an extension $L$ roughly given by  the speed of sound multiplied by the typical time $\Delta t$ taken by a mode to cross this region. Therefore, by using Eq.\ (\ref{entropy}), we can write  $\dot S=1/( 6 \Delta t)$. On the other hand,  the surface gravity can be interpreted as the inverse of the time taken by a mode to cross the near-horizon region. Therefore, we find that $\dot S=\kappa/6$, in line with \cite{giovanazzi}, up to a factor of two, that depends on considering bosons rather than fermions. Although there is no explicit time-dependence in the model, the Hawking mechanism still produces a steady flow of outgoing phonons at a rate that depends on $\kappa$. Therefore, the time derivative of $S$ should be interpreted as the production rate of the entropy associated to this flow.

%%%%%%%%%
\section{Dispersive case}
%%%%%%%%%
We now consider the dispersive case, and evaluate the contribution to the entropy given by high frequency modes. The two equations of the system (\ref{dispeqns}) can be easily decoupled in momentum space. By defining $\Psi^{\pm}(t,x)=\exp(i\omega t)[\phi(x)\pm\varphi(x)]$, we find that the Fourier transforms of Eqs.\ (\ref{dispeqns}) reads
\bea\non
\hat c^{2} \tilde \Psi^{+}(p)-\left[ {(\omega-pv)^{2}\over p^{2}}-{\hbar^{2}p^{2}\over 4m^{2}} \right]\tilde  \Psi^{+}(p)&=&0\ ,\\
\tilde \Psi^{-}(p)-{2m(\omega-pv)\over \hbar p^{2}}\tilde \Psi^{+}(p)&=&0\ .
\eea
The first of these equations can be solved with the WKB method in the near-horizon approximation $\hat c=v+i\kappa \partial_{p}$. At the leading order, we find
\bea
\tilde f(p)={v\hbar\over \kappa}\left(1+\sqrt{\left(1-{\omega \over vp}\right)^{2}-{\xi_{0}^{2}p^{2}\over 4}}\right)\ ,
\eea
where $\xi_{0}=\xi(x=0)$ is the healing length at the horizon. 
%The positive sign of the square root is chosen in order to recover the non-dispersive result (\ref{ndsoln}) in the limit $p\xi_{0}\rightarrow 0$. 
The number of modes is
\bea\label{numbermodesdisp}
\tilde n(\omega)={v\over \pi\xi_{0}\kappa}\int_{x_{\rm min}}^{x_{\rm max}}dx\left[1+\sqrt{\left(1-{a\over x}\right)^{2}-{x^{2}\over 4}}\right],
\eea
where $x=p\xi_{0}$ and $a=\omega\xi_{0}/v$. The integration boundaries are fixed by the positivity of $p$ and of the argument of the square root, namely $0<x<-1+\sqrt{1+2a} \quad\cup\quad 1-\sqrt{1-2a}<x<1+\sqrt{1-2a}$. However,  as the wavelengths of the physically realistic modes, given by $v/\omega$, must be much larger than $\xi_{0}$, we see that $a\ll 1$. Thus, we can simplify the above integral by expanding the integrand function and the upper integration limit as
\bea
\tilde n(\omega)\simeq{v\over \pi\xi_{0}\kappa}\int_{\epsilon}^{a}dx\left({a\over x}-{x^{3}\over 8a}\right),
\eea
where $\epsilon = \xi_{0} p_{\rm min}\ll a$ is set to cope with the same logarithmic divergence encountered in the non-dispersive case. With the help of Eqs.\ (\ref{freeen}), we calculate the entropy in the form $S=S_{\rm lead}+S_{\rm corr}$ where
\bea
S_{\rm lead}&=&{1\over 4}-{\gamma\over 6}+{\zeta (1,2)\over \pi^{2}}+{1\over 6}\ln \left( \kappa\over 2\pi vp_{\rm min}\right)\ ,\\
S_{\rm corr}&=&-{1\over 960} {\xi_{0}^{2}\kappa^{2}\over v^{2}}\ .
\eea
In these expressions $\gamma$ is the Euler constant and $\zeta(1,2)$ is the first derivative of the $\zeta$-function evaluated at 2. To obtain this result, we assumed that the inverse temperature of the Hawking radiation is  $\beta=2\pi/(\hbar\kappa)$, i.e.\  it is not affected by dispersion \cite{cimento,analyticBEC}.

%%%%%%%%%
\section{Discussion}
%%%%%%%%%

We first note that the correction $S_{\rm corr}$ to the entropy is set by the square of the size of the near-horizon region $L\simeq v/\kappa$ measured in units of the healing length. Remarkably, this term is very similar to the one found in the case of  the one dimensional spin-1/2 Heisenberg XX chain in a magnetic field \cite{calabrese}. 

Another important observation is that the leading term $S_{\rm lead}$ is no longer completely arbitrary as in the hydrodynamic case. In fact, dispersion affects the integration boundaries in Eq.\ (\ref{numbermodesdisp}), which are no longer put by hands, but are fixed in the ultraviolet by the system itself, as the dispersion relation in this system is fixed by the properties of the BEC and can be found to be expressed by the function \cite{mayo-fabbri-rinaldi}
\bea
(\omega-vp)^{2}=c^{2}\left(p^{2}+{\xi^{2}p^{4}\over 4}\right)\ .
\eea
As a result, the leading term turns out to be a numerical constant plus a logarithmic correction that depends on the infrared regulator. Physically, the argument of the logarithm  $\kappa/(2\pi v p_{\rm min})$ can be written as the ratio of the maximum wavelength allowed in the system $\lambda_{\rm max}=1/p_{\rm min}$ and the size of the box, i.e. of the near-horizon region, $L$.  It is reasonable to expect that, by taking in account the dependence of $\theta_{1}$ upon transverse directions (while keeping the potential a function of $x$ only) one can provide for an effective mass able to act as an infrared regulator. This issue will be investigated in a future paper.

\begin{figure}[h]
\centering
\includegraphics[width=8.5cm]{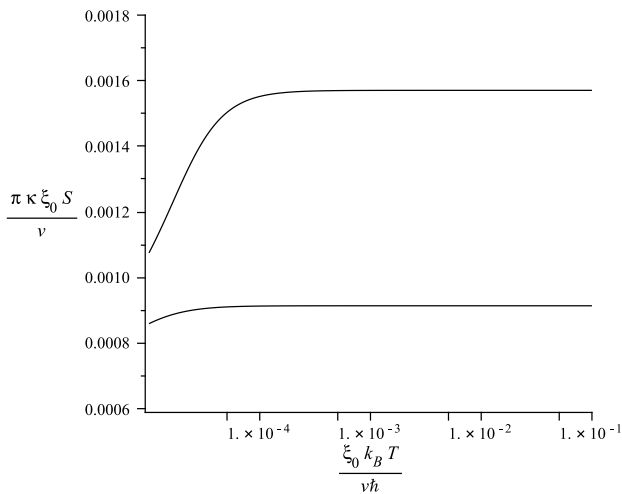} 
        \caption{Normalized entropy versus the temperature for small temperatures. The bottom curve is obtained by subtracting the function $a/x$ from the integrand of Eq.\ (\ref{numbermodesdisp}) and setting $x_{\rm min}=0$. The top curve is obtained by setting $x_{\rm min}=10^{-12}$. }
	\label{plot2}
\end{figure}

To further check our results, we compute the entropy numerically, and we cope with the infrared divergence in two ways. In the first, we set the value of $x_{\rm min}=10^{-12}$ in the integral (\ref{numbermodesdisp}). In the second, we subtract to the integrand the function $a/x$ and we let $x_{\rm min}=0$. The upper integration limit is set at $x_{\rm max}=10^{-4}$ in both cases. The two results are plotted in Fig.\ (\ref{plot2}), and we see that the entropy is in fact constant in terms of the normalized temperature $\xi_{0} k_{\rm B} T/(v\hbar) \simeq  \xi_{0}/( 2\pi L)$. The different numeric values of the plateaux depend only on the choice of the infrared regularization.  If one considers typical experimental values for Rubidium atoms, like $v= 4\times 10^{-3}\, m/s$, $\xi_{0}=2\times 10^{-7}\, m$, $\kappa=2,7 \times 10^{3}\, Hz$, the Hawking temperature is  of the order of few $nK$ \cite{balb-carus-fabbri}, which corresponds to values on the horizontal axis around 0.05. We see that the entropy is constant in a large range containing this value. This is expected as the first order correction, which can be written as $S_{\rm corr}\simeq -(\pi\xi_{0}k_{\rm B} T/4v\hbar)^{2}/15$, is very small for these values.

In summary, we have verified that the scaling behavior of the entropy in the hydrodynamic limit is the same as the one predicted by conformal field theory, by using a method inspired by the brick wall model for astrophysical black holes. When dispersion is taken into account, we found a correction that is similar to the one calculated for the entanglement entropy of certain spin-chain systems. Also, the leading term appear to be a constant determined uniquely by the infrared cut-off. These elements, although not a rigorous proof, strongly support the interpretation of  the brick wall entropy as due to the entanglement of the phonon pairs created via the Hawking mechanism.

We thank R.\ Balbinot, P.\ Calabrese, and S.\ Giovanazzi for stimulating discussions during the workshop ``New trends in the physics of the quantum vacuum: from condensed matter, to gravitation and cosmology '' organized by the ECT* of Trento, Italy.

%%%%%%%%%%%%%%%%%%%%%%%%%%
%%%%%%%%%%%%%%%%%%%%%%%%%%

\end{document}